\documentclass[a4paper,final]{article}

\usepackage{jheppub}
\usepackage{amsmath,amssymb,amsthm,amscd}
\usepackage{graphicx}
\usepackage{subcaption}

\author{Santiago Codesido S\'anchez}
\title{On the resummation of the Lee-Yang edge singularity coupled to gravity}
\abstract{
	We study the Borel-Pad\'e resummation of the asymptotic series for the string equation of the Lee-Yang edge singularity. Numerical methods
	are provided to compute a high accuracy exact solution. We find the resummation matches the numerical integration
	without need for further non-perturbative corrections. }

\begin{document}
	\maketitle
	
	\tableofcontents
	
	\section{Introduction}

As it is well known, many of the usual perturbative series used in physical problems are only asymptotic in nature. 
A standard tool to make sense of these formal expansions is Borel resummation \cite{cal-mey-rib-sur-jen}. However, it is often the case that one has to add 
additional ``non-perturbative" sectors to the original perturbative series, in order to reconstruct the exact answer by resummation. This leads to the theory of 
transseries and resurgent asymptotics. For example, in the theory of ODEs, the asymptotic expansion near an irregular singular point 
can be extended to a transseries which might then be used to reconstruct the exact solution (see \cite{mar} for a comprehensive review of these and other examples).

The genus expansion of string theory is no exception to this divergent behaviour \cite{gro-per},
but in this case it is not clear what is the exact, non-perturbative answer behind the perturbative series. 
In some examples, one can use non-trivial dualities, such as the AdS/CFT correspondence, to provide an exact answer from the dual side. 
This opens up the possibility of comparing the Borel resummation of the divergent string series with an
exact definition.

This comparison was done for the free energy on the three sphere of ABJM theory in \cite{gra-mar-zak}, which can be computed exactly by localization techniques. 
The genus expansion of string theory is simply the 
$1/N$ expansion of this free energy. It turns out that this expansion is Borel summable, i.e. there are no obstructions to the Borel resummation of the series. 
However, and in contrast to what happens in many simple models in Quantum Mechanics, 
this resummation does not agree with the exact result. Of course, there is no obvious \textit{a priori} reason why there should be such an agreement. 
The difference between the two results was understood in terms of complex instantons, which show up as complex poles in the Borel integration plane.
A closely related example was recently studied in the context of topological string theory on toric Calabi--Yau threefolds. In these models, one 
can provide a non-perturbative definition of the topological string free energy by using spectral theory \cite{gra-hat-mar} (see \cite{mar2} for a review). 
Once again, Borel resummation of the asymptotic series differs from this exact result. 
However, the perturbative solution can be upgraded to a transseries including instanton corrections \cite{cou-ede-sch-von}. 
By a systematic Borel resummation of the full transseries \cite{cou-mar-sch}, the exact answer can be indeed reproduced.

These non-trivial facts led us to review earlier models of strings with an exact, non-perturbative definition. We consider in this paper 
minimal models coupled to 2D gravity \cite{dou-she,bre-kaz,gro-mig}, which can be defined through the double scaling limit
of a theory of randomly triangulated surfaces \cite{kaz-kos-mig}. One can even write an ODE 
for the partition function of these non-critical string models, depending on a parameter essentially equivalent to the string coupling constant. 
This ODE is called in this context the string equation of the model. 
The asymptotic expansion of the solution to this ODE 
corresponds the genus expansion of the corresponding string. 

The existence of the string equation is not enough to define the theory non-perturbatively (see \cite{dif-gin-zin} for a review), since one has to 
provide in addition appropriate boundary conditions. In the case of the Lee-Yang edge singularity coupled to gravity, such a boundary condition can be found, 
and one can use it to calculate the free energy 
of the non-critical string exactly, by numerical integration of the string equation \cite{bre-mar-par}.
The corresponding asymptotic series is known to be Borel summable, and in \cite{dif-gin-zin} it is asked whether its Borel resummation reproduces the true solution to the string equation. 
In view of the recent results in \cite{gra-mar-zak,cou-mar-sch}, this is not obviously the case. We will show in this paper, by a detailed numerical comparison of the resummation and the numerical 
integration of the string equation, that the solutions seem to be equal, at least with a very high numerical accuracy. 
This is the case \textit{despite} the presence of complex Borel poles like the ones identified in \cite{gra-mar-zak} as a source for the corrections to the Borel resummation.

This paper is organized as follows. We will first briefly review the analytic structure of the solutions and their large order behaviour,
and then we will compare the Borel resummation with the exact numerical integration.
	
	\section{The string equation of the Lee-Yang edge singularity}

\subsection{Perturbative solution}
The string equation of the Lee-Yang edge singularity itself is given, as in \cite{bre-mar-par}, by
\begin{equation}
	P[f,x] = f^3 + \frac{1}{10} f^{(4)}+f f'' + \frac{1}{2} \left(f'\right)^2 - x = 0.
	\label{eq:ly:LYeq}
\end{equation}
The physical conditions we must impose to the solution, following \cite{bre-mar-par}, are absence of poles (which would mean zeros in the partition
function) and most importantly, an asymptotic behaviour that gives the right large $N$ limit for the string picture. It is
\begin{equation}
	f\left(x\right) \underset{|x|\to \infty}{=} \textrm{sign}\left(x\right) |x|^{1/3} + o\left(|x|^{1/3}\right).
\end{equation}
In \cite{bre-mar-par} a numerical plot of a numerical approximation to the pole-free solution is provided. For negative values of $x$, the solution oscillates. This is an
Airy-like symptom of the Stokes phenomenon (see \cite{mar}). Essentially, complex exponentially small corrections take over, and their interference produces
the oscillation.
This will set our focus on the $x>0$ region. There, the series expansion at infinity is resummable. Since no oscillations are present there is no \textit{a priori} need
to have exponentially small corrections to the Borel sum. The question is: does the resummation, with no
further corrections, agree with the numerical answer?

We will begin by computing this resummation.
The limit in which we implement the asymptotic conditions is $x\to\infty$. For convenience, we
redefine $f(x)=x^{1/3}g(t)$ and $x=t^{-3/7}$. In this variable, (\ref{eq:ly:LYeq}) is solved by the (formal) series
\begin{equation}
	g(t) = \sum_{k=0}^\infty \mu_k t^{k}.
	\label{eq:ly:ansatz}
\end{equation}
The coefficients can be recovered algebraically order by order, and we find
\begin{gather}
	\begin{gathered}
	\mu_0=1,\:\:\mu_1=\frac{1}{18},\:\:\mu_2=-\frac{7}{108},\:\:\mu_3=\frac{4199}{17496},\:\:\mu_4=-\frac{409297}{262440},\:\:\mu_5=\frac{101108329}{9447840},\\
	\:\:\mu_6=\frac{25947984239}{191318760},\:\:\mu_7=-\frac{3760665121759}{204073344},\:\:\mu_8=\frac{1158425083469857567}{826497043200},\dots
	\end{gathered}
\end{gather}
It is easy to generate them with computer aid at much higher orders, as will be required later for numerics.

\subsection{Borel transform}
\begin{figure}[h]
	\caption{Growth of coefficients}
	\begin{center}
		\begin{subfigure}[b]{0.45\linewidth}
			\caption{$\left|\mu_k\right|$}
			\includegraphics[width=\linewidth]{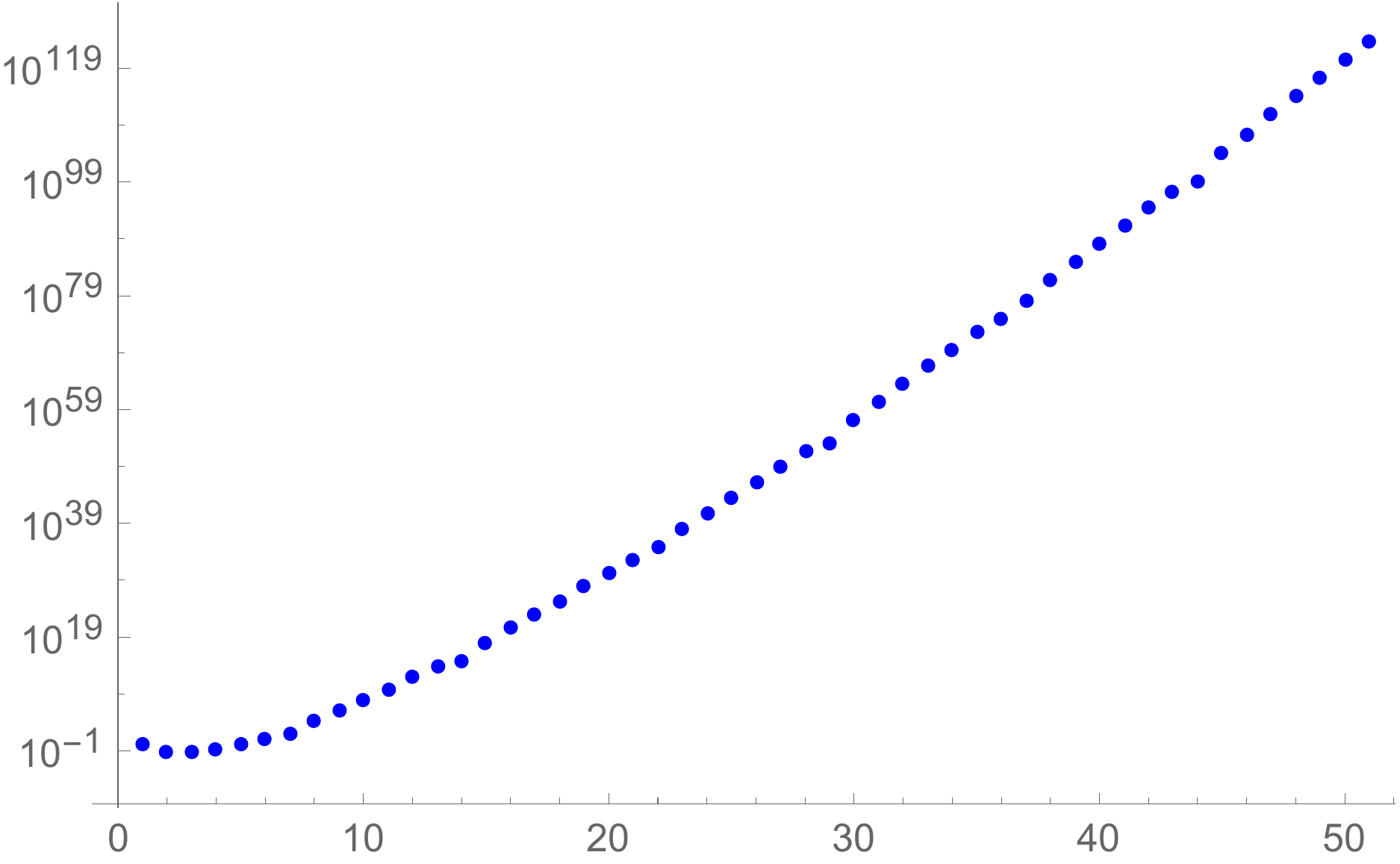}
		\end{subfigure}
		\begin{subfigure}[b]{0.45\linewidth}
			\caption{$\textrm{sign}\left(\mu_k\right)$}
			\includegraphics[width=\linewidth]{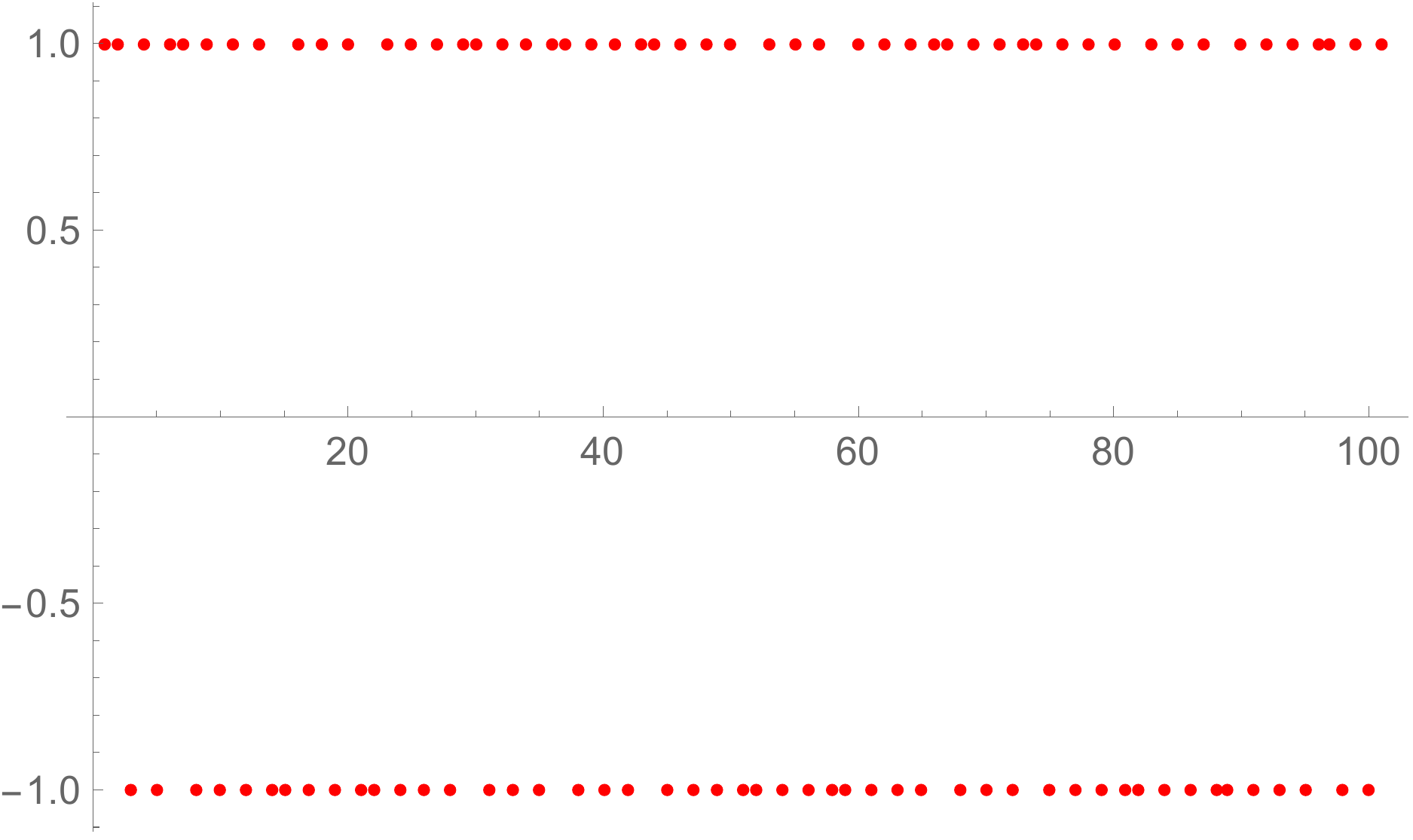}
		\end{subfigure}
	\end{center}
	\label{fig:ly:growth}
\end{figure}

That (\ref{eq:ly:ansatz}) is only an asymptotic series is hinted at already by the first few coefficients. The
divergence can also be seen in figure \ref{fig:ly:growth}. Our objective now is to make sense of this formal, divergent, series by using Borel resummation.

The idea of the Borel transform is to fix a $(\beta k)!$ divergence in a formal power series by defining the (now converging) series
\begin{equation}
	\left(B g\right)(t) = \sum_{k=0}^\infty \frac{\mu_k}{(\beta k)!} t^{k} := \sum_{k=0}^\infty \tilde{\mu}_k t^k.
\end{equation}
and then reintroducing the divergence
 ``inside'' of the sum by commuting it with the integral form of the gamma function. This defines the Borel resummation of $f\left(x\right)$,
\begin{equation}
	F(x) = x^{\frac{1}{3}+\frac{7}{3\beta}}\int_0^{\infty} e^{-x^{\frac{7}{3\beta}}s} \left(B g\right)\left(s^\beta\right) \: \mathrm{d}s.
	\label{eq:ly:borelSum}
\end{equation}
The motivation for this definition is that, should everything converge, the Borel resummation gives the actual value of the original series.

The particular rate of divergence that we have, as is the usual case with string theory, is $\beta=2$. 
This asymptotic behaviour 
can also be found by looking at the first exponentially small correction as done in \cite{dif-gin-zin}.

\begin{figure}[h]
	\caption{Borel coefficients $\frac{|\mu_k|}{(2k)!}$}
	\centering
	\includegraphics[width=0.6\linewidth]{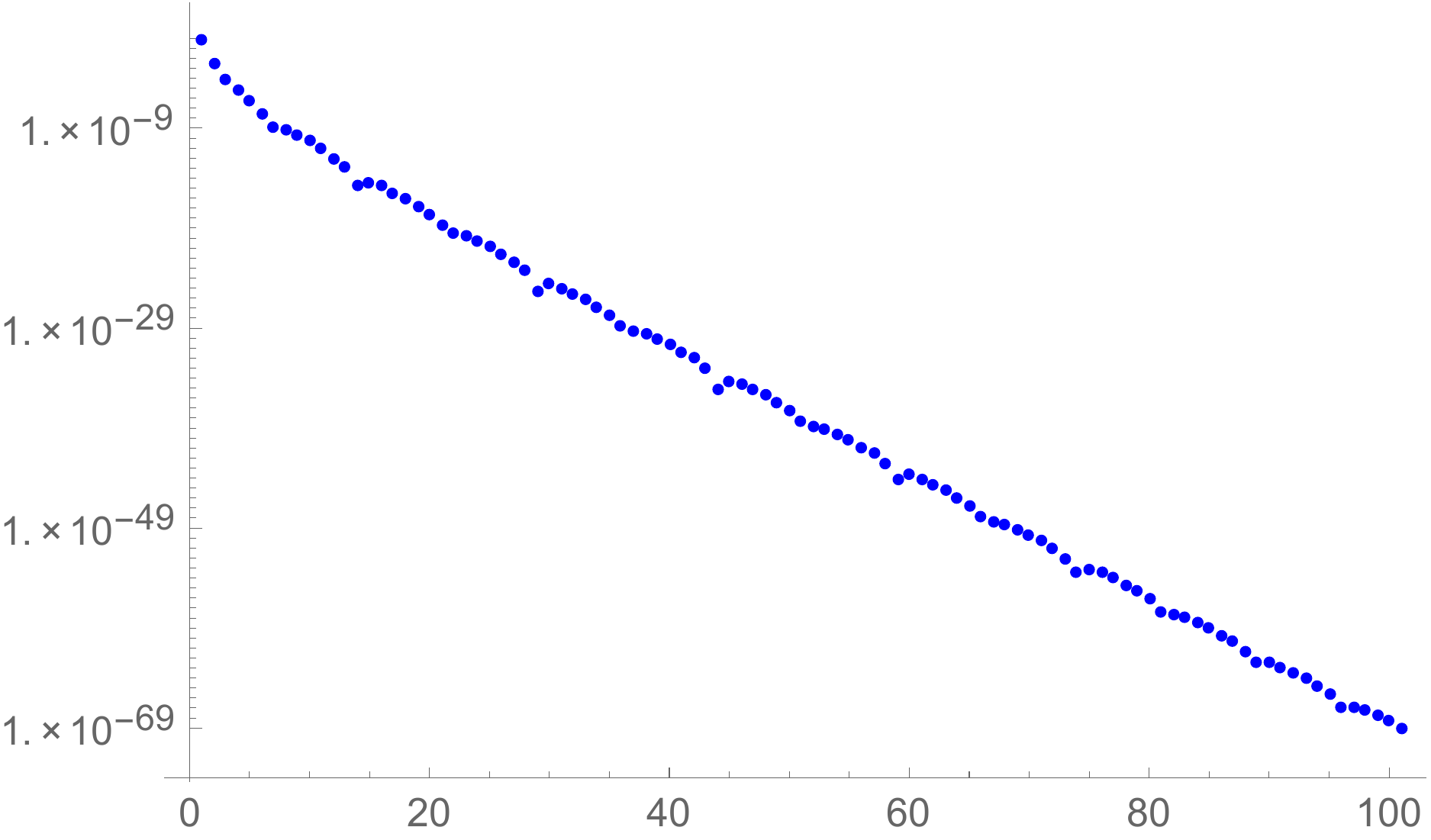}
	\label{fig:ly:borelGrowth}	
\end{figure}

In figure \ref{fig:ly:borelGrowth} we have
the coefficients of $\left(B g\right)$, which now have a behaviour similar to $|\tilde{\mu}_k| \sim A^k$, with a finite radius of convergence. Still,
to have a well defined integral in (\ref{eq:ly:borelSum}), we need the integration path in (\ref{eq:ly:borelSum}) to be free of obstructions.

Suppose we had
$\tilde{\mu}_k = A^{-k}$ with $A>0$. This can be directly resummed, and $\left(B g\right)$ would have a pole as $\left(t-A\right)^{-1}$, right in the middle
of the integration path. In our case,
like we saw in figure \ref{fig:ly:growth}, the
signs are alternating. In the simplified version $\tilde{\mu}_k = A^{-k}$,
it would happen when $A<0$ -- the pole would lie on the \textit{negative} real axis, leaving the path of integration free.

We will need the derivatives $F^{(m)}$ of (\ref{eq:ly:borelSum}) for our numerical analysis. Since for $x \neq 0$ the integrand of $F(x)$ and its derivatives are
continuous, there are no problems in commuting $\partial_x$ with the $s$-integral, and one gets expressions like
\begin{equation}
	F^{(1)}(x) = x^{\frac{1}{3}+\frac{7}{3\beta}}\int_0^{\infty} \frac{\beta -7 s x^{\frac{7}{3\beta} }+7}{3 \beta  x}
	e^{-x^{\frac{7}{3\beta}}s} \left(B g\right)\left(s^\beta\right) \: \mathrm{d}s.
	\label{eq:ly:borelSumFirstDerivative}
\end{equation}	

\subsection{Pad\'e approximant and Borel plane}
\begin{figure}[h]
	\centering
	\caption{Borel-Pad\'e $t$-plane}
	\begin{subfigure}[t]{0.9\linewidth}
		\caption{Poles (left) and zeros (right) of $P^{(20/20)}$}
		\includegraphics[width=\linewidth]{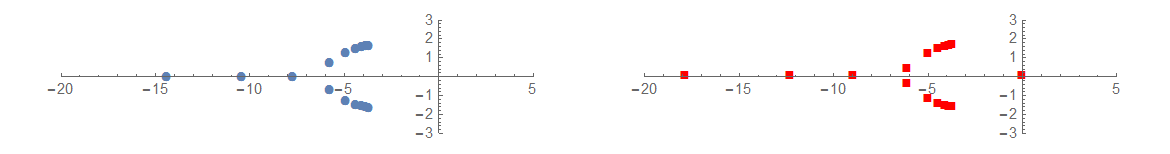}
	\end{subfigure}
	\begin{subfigure}[b]{0.9\linewidth}
		\caption{Poles (left) and zeros (right) of  $P^{(70/70)}$}
		\includegraphics[width=\linewidth]{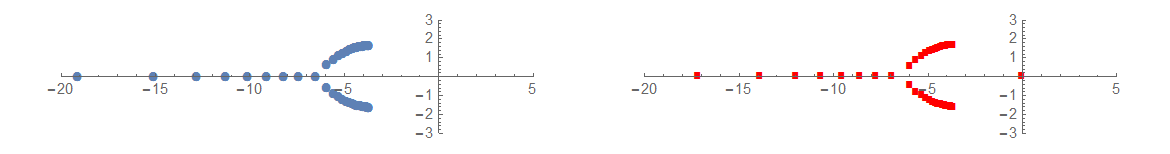}
	\end{subfigure}
	\label{fig:ly:borelPlane}
\end{figure}

In the previous section we considered the simple case of a Borel transform behaving exactly like $\tilde{\mu}_k = A^{-k}$.
In general, one cannot find such a closed expression for $\left(B g\right)$ -- let alone when we do not even have a closed form for the $\mu_k$, and
are generating them recursively. However,
 a good (and integrable) substitute to an exact analytical continuation is given
by the Pad\'e approximant, as detailed in \cite{cal-mey-rib-sur-jen}. In short,
it is the rational function whose Taylor expansion agrees with the original series at a given order.
We denote by $P^{(n/n)}$ the diagonal approximant matching the series $\left( B g\right)$ at order $O(t^{2n+1})$. That is, the one with equal degree in numerator and denominator.
For Borel-Pad\'e resummation, it is the one\footnote{
	To work with derivatives such as (\ref{eq:ly:borelSumFirstDerivative}) it is more convenient to use shifted Pad\'e approximants such as $P^{(n-1/n+1)}$, 
	to compensate the $s$ factors introduced by the $x$-derivative, so that we still get good convergence in the numerical integration
	for the relatively smaller values of $x$. Of course, the picture we get of the Borel plane for them is essentially the same as in figure (\ref{fig:ly:borelPlane}).}
that usually gives the best numerical convergence with $n$.
	
As we see in figure \ref{fig:ly:borelPlane}, where we represent
the poles and zeros of $P^{(n/n)}\left(t\right)$ in the Borel $\mathbb{C}$-plane,
there is a nice structure that remains stable as we improve the approximation.
In particular, a branch cut develops
along the negative axis, leaving no obstruction for the integral along $x>0$.

\begin{figure}[h]
	\caption{Borel-Pad\'e $t$-plane and action $A^2$}
	\centering
	\includegraphics[width=0.6\linewidth]{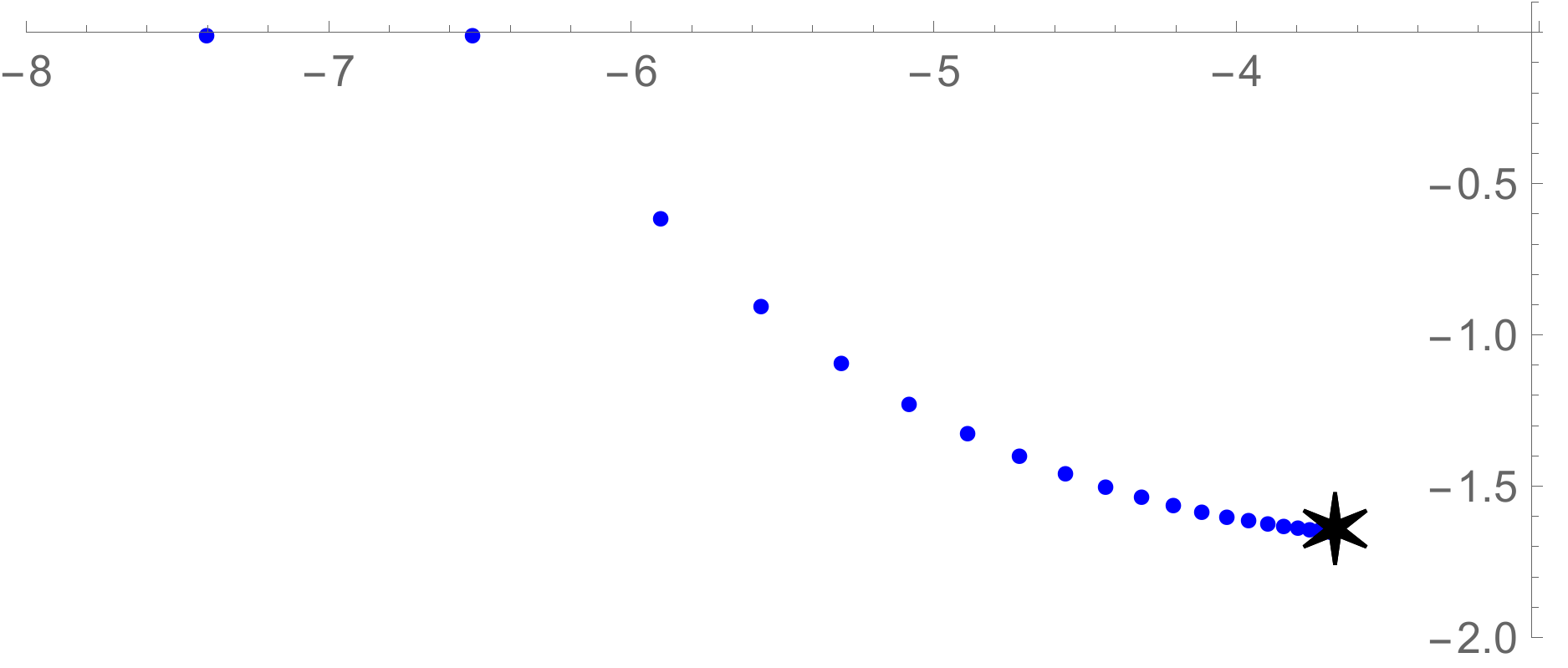}
	\label{fig:ly:action}
\end{figure}

Let us remark that the closest poles to the origin are located precisely at the values of the action $A^2$ given by the first exponential correction.
This is well known, and \cite{eyn-zjt} provides a general way to compute the action for an arbitrary $(p,q)$ model.
A detailed calculation for our particular case is given in \cite{dif-gin-zin}. For illustrative purposes consider simply the ansatz
\begin{equation}
	g(t) = g_{(0)}(t) + g_{(1)}(t) e^{-\frac{A}{\sqrt{t}}} + \dots
\end{equation}
where $g_{(0)}$ is the perturbative solution of (\ref{eq:ly:ansatz}), and $g_{(1)}$ the power series
for the exponentially suppressed part -- the \textit{one-instanton transseries}. Inserting it into (\ref{eq:ly:LYeq}), the first-order condition is
\begin{equation}
	\frac{2401 A^4}{12960} +\frac{49 A^2}{36}+3 = 0.
\end{equation}
The solutions\footnote{
	$A$ itself is of course chosen by having positive real part, so the instanton correction is small.}
to this are $A^2 = -\frac{36}{49} \left(5 \pm i \sqrt{5}\right) \simeq -3.67 \pm 1.64 \: i$,
plotted as a star in figure \ref{fig:ly:action}, matching the two closest poles to the origin in the Borel plane.

The natural variable for the transseries is, however, $\sqrt{t}$. In that plane, 
easily obtained by unfolding the one in figure (\ref{fig:ly:borelPlane}),
there are two branch cuts along the imaginary axis,
and we actually have four actions, two of
which have positive real part. As discussed in \cite{cou-mar-sch}, the transseries associated to such actions
is related to the difference between Borel resummation and their exact answer. Remarkably, we will see in the next section that
despite the presence of positive real part poles, we will not need any transseries correction.

	\section{Numerical analysis}

\subsection{Convergence of the resummation}
We integrate (\ref{eq:ly:borelSum}) numerically. We will truncate $F(x)$ at order $n$ by using the $P^{(n/n)}$ approximant in (\ref{eq:ly:borelSum}). This gives
an approximation $F_n(x)$ to $F(x)$.
Assuming that the error made by truncating decreases with $n$, we can estimate the relative uncertainty due to truncation by

\begin{equation}
	\epsilon[F_n](x) := \frac{\Delta[F_n](x)}{\left|F_n(x)\right|} := \left| \frac{F_{n}(x)-F_{n-1}(x)}{F_n(x)} \right| .
	\label{eq:num:errorDef}
\end{equation}

\begin{figure}[h]
	\centering
	\caption{Numerical Borel resummation, as a function of $x$}
	\begin{subfigure}[b]{0.45\linewidth}
		\includegraphics[width=\linewidth]{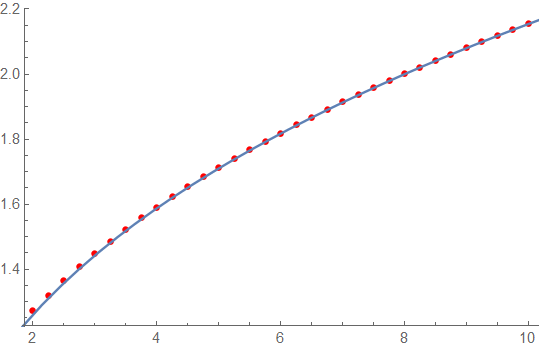}
		\caption{Borel function evaluations,\\ $F_{70}(x)$ vs. $x^{1/3}$}
	\end{subfigure}
	\begin{subfigure}[b]{0.45\linewidth}
		\includegraphics[width=\linewidth]{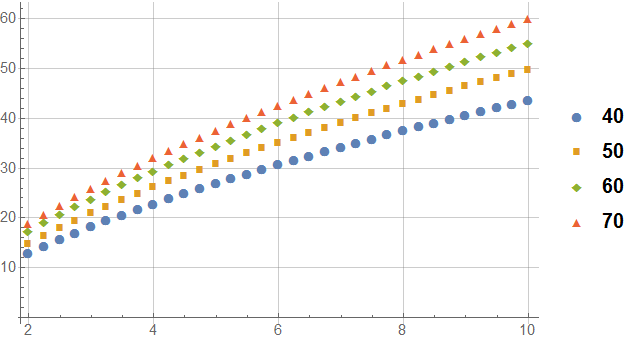}
		\caption{Stable digits of $F_n(x)$, \\ given by $-\log_{10} \epsilon[F_n](x)$}
	\end{subfigure}
	\label{fig:num:borelFunction}
\end{figure}

In figure \ref{fig:num:borelFunction} we plot the calculated values of (\ref{eq:ly:borelSum}), and the asymptotic function $x^{1/3}$ -- the difference
becomes only obvious at $x\sim 1$.
Also, for different truncations,
the number of reliable digits is given. These are computed by looking at how many digits remain stable from one truncation to the previous, 
which is in essence the logarithm of the relative error.
We can easily see the convergence as we increase $n$, with stable digits ranging between 20 to 60, depending on the value of $x$. One finds similar
plots for the first four derivatives $F^{(m)}_n(x)$. This validates, in retrospect, the assumption that (\ref{eq:num:errorDef}) is a good estimate for the truncation error.

\subsection{First approach: error propagation}
We will try to answer whether the resummation solves (\ref{eq:ly:LYeq})
without requiring direct knowledge of the exact (numerically integrated) solution.
We recast the equation into a normalized form $Q [f,x] = P[f,x]/x = 0$,
so that it measures in a homogeneous way the failure of $F_n$ (and its derivatives) to be a solution of (\ref{eq:ly:LYeq}),
\begin{equation}
	E_n(x) = \left|Q[F_n,x]\right|.
	\label{eq:LYEqfailure}
\end{equation}
The error $E_n$ should be zero in the $n\to \infty$ limit. That is, if $F(x)$ --the Borel sum-- is really a solution of (\ref{eq:ly:LYeq}).
The question is then whether the deviations thereof at finite $n$ are only due to $F_n$ itself being only an approximation of the full $F$, or rather to $F$ being close to
but not a solution of (\ref{eq:ly:LYeq}). To check this, we can propagate the uncertainty of $F_n(x)$ in (\ref{eq:num:errorDef}). For the sake of simplicity we use the approximation of the small-error
propagation formula,
\begin{equation}
	\Delta\left[ Q[F_n,x] \right] = \sqrt{\sum\limits_{i=0}^4\left(\frac{\partial Q}{\partial f^{(i)}}\left[F_n,x\right]\right)^2 \left(\Delta\left[F^{(i)}_n\right]\right)^2}.
	\label{eq:errorPropagation}
\end{equation}
\begin{figure}[h]
	\centering
	\includegraphics[width=0.5\linewidth]{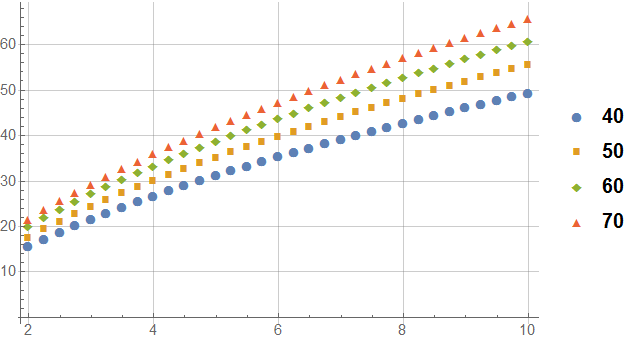}
	\caption{$-\log_{10} \Delta\left[ Q[F_n,x] \right]$}
	\label{fig:num:propagatedEquationError}
\end{figure}
Figure \ref{fig:num:propagatedEquationError} gives the expected correct digits from the uncertainty propagation. The plot for the error $E_n$ is fundamentally the same,
and we empirically find (with $n$ up to 70) that
\begin{equation}
	0 < \log_{10} \frac{\Delta\left[ Q[F_n,x] \right]}{E_n(x)} < 1,
\end{equation}
that is to say, the failure of $F_n(x)$ to be a solution is just somewhat smaller
than what one would expect from simply propagating the truncation errors. This not surprising since
(\ref{eq:errorPropagation}) can slightly overestimate the error due to correlations. Therefore we cannot say $F(x)$ does not solve (\ref{eq:ly:LYeq}) up to the numerical accuracy.
In other words, it is a \textit{sufficiently} good solution 

Finally, figure \ref{fig:num:propagatedEquationError} also shows this error decreases steadily with the Pad\'e order, across all values of $x$. This is
important to rule out the possibility that simply the trivial, leading part is responsible for the matching: the Borel sum is in fact \textit{necessary}
to tend to the correct answer. We will test this in more detail in the next section.

\subsection{Second approach: numerical integration}
Although it is useful being able to face the problem with a minimal amount of information, a direct comparison is of course a stronger test.
In order to find the exact numerical solution, we first turn our attention to how would it be possible to implement 
the boundary conditions of the problem. The series being asymptotical in nature,
our best bet is to take some \textit{optimal truncation} as a starting guess in a region where it allows for reasonable precision -- at least, of the order
of the Borel resummations we were considering.
\begin{figure}[h]
	\caption{$ \mu_k \: x^{\frac{1-7k}{3}} $, as a function of $k$}
	\centering
	\begin{subfigure}[b]{0.45\linewidth}
		\includegraphics[width=\linewidth]{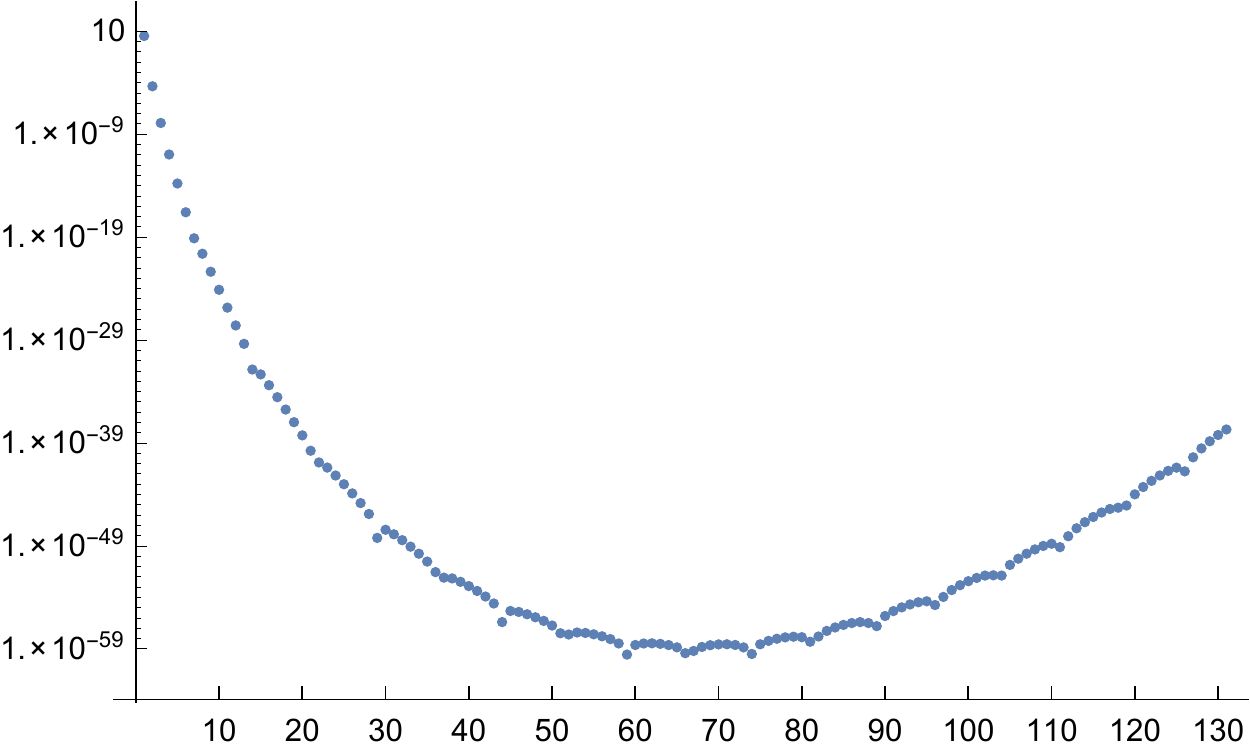}
		\caption{$x=36$}
	\end{subfigure}
	\begin{subfigure}[b]{0.45\linewidth}
		\includegraphics[width=\linewidth]{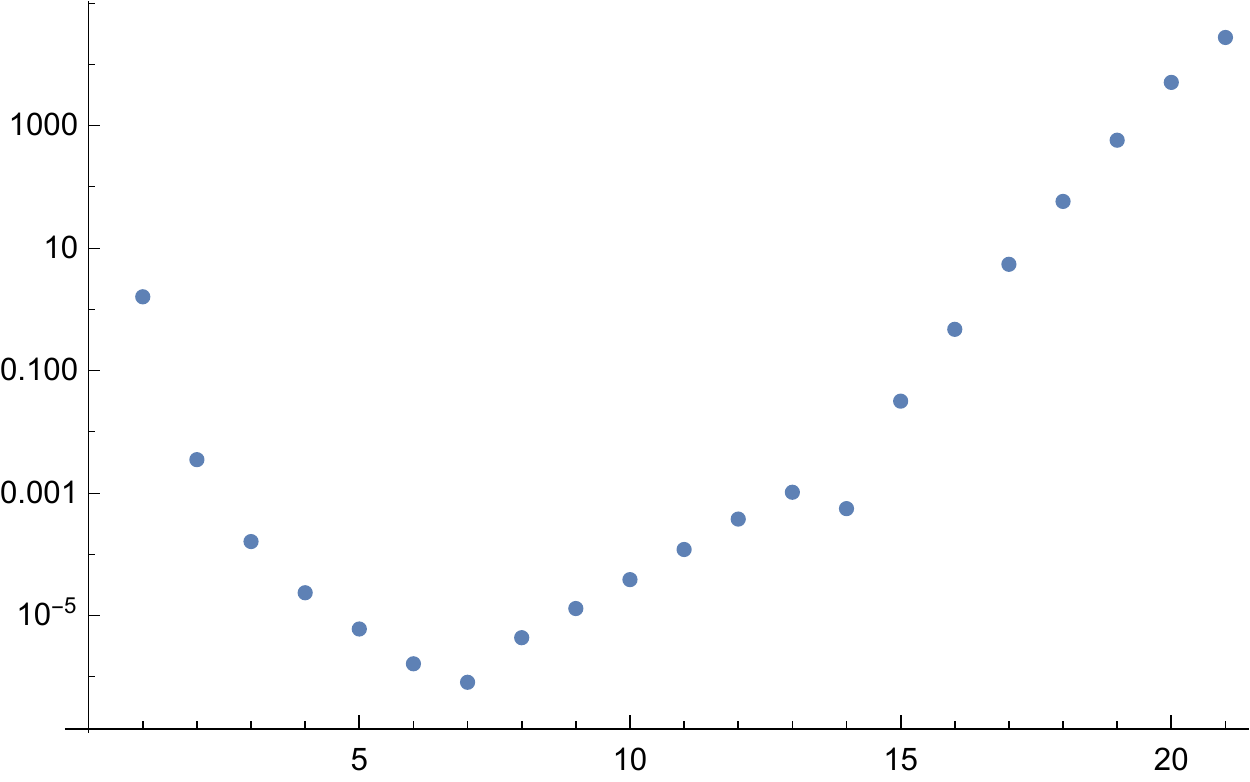}
		\caption{$x=4$}
	\end{subfigure}
	\label{fig:num:optimalTruncation}
\end{figure}

Figure \ref{fig:num:optimalTruncation} shows how much optimal truncation breaks down at small values of $x$.
Including terms up to Pad\'e order $n=65$ (or Taylor order $130$) for $x=36$, we get to precisions of $\sim 60$ digits. To give an example,
\begin{small}
\begin{align}
	\begin{split}
	\left. f(36)\right|_{n=65} &= 3.30197010415890744173610022362421011512463564025031823619490617\dots \\
	\left. f(36)\right|_{n=64} &= 3.30197010415890744173610022362421011512463564025031823619490258\dots
	\end{split}
	\label{eq:num:asymptoticGuess}
\end{align}
\end{small}
The derivatives, computed as in (\ref{eq:ly:borelSumFirstDerivative}), have the same amount of stable digits. For the sake of comparison,
if we try at $x=4$, the ``optimal'' truncation now happens at order 6, with only 6 reliable digits.
 
The standard procedure to backtrack along the $x$ axis would now be to use a finite step numerical algorithm. However, the Lee-Yang
equation is particularly troublesome, displaying stiffness problems that cannot be consistently avoided even by Runge-Kutta integrators. This is made
worse by the kind of precision we require to meaningfully compare with Borel-Pad\'e.

We will resort to a method similar to what is described in \cite{pra-spo}, in spirit a finite step method of arbitrarily high order.
Given a set of ``initial conditions'' $\left\{f^{(n)}(x_i),n=0\dots 3 \right\}$, one builds the power series expansion of $f(x)$ around $x_i$,
\begin{equation}
	f(x) = \sum_{k=0}^\infty \nu_k(x_i) \left(x-x_i\right)^k,
\end{equation}
and find the rest of the $\nu_k(x_i)$ recursively with (\ref{eq:ly:LYeq}), by requiring
\begin{equation}
	P\left[\sum_{k=0}^N \nu_k(x_i) \left(x-x_i\right)^k,x\right] = 0 + O\left(x^{N+1}\right).
	\label{eq:num:nuCondition}
\end{equation}
At finite $x_i$ the solution is regular and has a finite radius of convergence of (\ref{eq:num:nuCondition}).
We can use that to compute the first three coefficients at the next step
\begin{equation}
	\nu_k(x_{i+1}) = \frac{1}{k!} \sum_{l=k}^{N} \nu_l(x_i) \frac{l!}{(l-k)!} \left(x_{i+1}-x_i\right)^{l-k} + O(\Delta x^{N+1}), \:\:\: k=0\dots 3,
	\label{eq:num:nuRecursion}
\end{equation}
calculate the rest $k=4\dots N$ with condition (\ref{eq:num:nuCondition}), and repeat.

This procedure allows one to take step sizes only limited by the radius of convergence in (\ref{eq:num:nuRecursion}). Notice that precision can be arbitrarily increased
by the order $N$ of the approximation, and we can take relatively big ``jumps'' that avoid the decreasing step size problem inherent to stiff equations.
It should be remarked, however, that numerical rounding errors accumulate fast when doing the recursion from (\ref{eq:num:nuCondition}) and
the propagation (\ref{eq:num:nuRecursion}) -- very high numerical working precisions
are needed to avoid them.

In the appendix we provide the values for both the Borel resummation at order $n_{\mathrm{Pade}}=60$, and
for the numeric integration at order $N=70$, calculated from the asymptotic guess (\ref{eq:num:asymptoticGuess}) with order $n_{\mathrm{guess}}=65$.
For the Borel sum, the stable digits are determined by contrast with the order $n_{\mathrm{Pade}}=59$ approximant. For the numeric integration,
they are given by comparison with a truncation at $N=69$ \textit{and} an initial guess at $n_{\mathrm{guess}}=64$. 

\begin{figure}[h]
	\centering
	\caption{Digit comparison (Borel vs. integration), as a function of $x$}
	\includegraphics[width=0.95\linewidth]{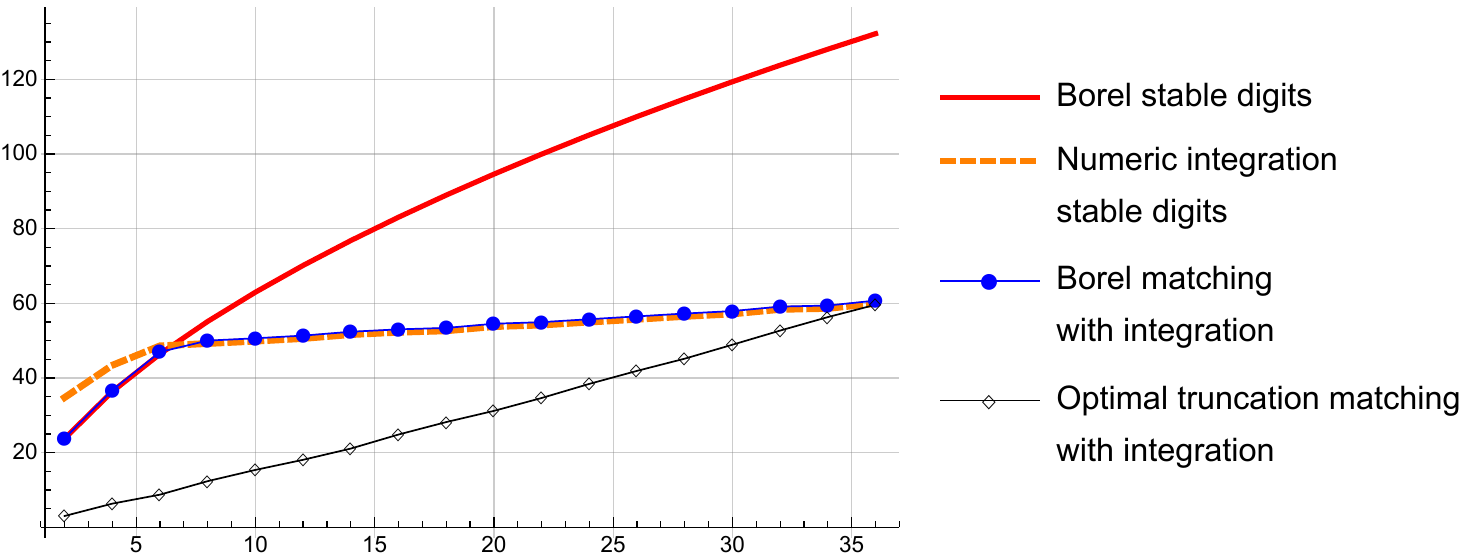}
	\label{fig:borelVsIntegration}
\end{figure}

In figure \ref{fig:borelVsIntegration} we have plotted the stable digits of each of the solutions, for values $x=2,4,\dots,36$.
The numeric integration is mostly limited by the error
of the initial guess. At very small values of $x$ ($\sim 5$) it starts to slowly break down due to an
insufficient truncation order $N$, but by then the Borel sum is performing worse. Of course,
$N$ was chosen taking this into account.
One could in principle go to higher orders in the asymptotic guess by using higher starting points in $x$. This implies a 
tradeoff with the numerical integration, and we would need even higher $N$ (and thus working precision) to be able to reach accurately the low $x$ region. 
In any case, around $x=36$ the convergence of Borel-Pad\'e already outperforms greatly the precision of the numerical integration. After all,
the initial guess (and the initial error of the integration) is given precisely by optimal truncation, 
and comparing it with the resummation does not say anything useful. The interesting section of the plot lies at the smaller values of $x$, where 
optimal truncation becomes a really bad approximation to the numerical result.

In general, the coincident digits between Borel and integration are just slightly better than
expected by the corresponding error of
the worse of the two. Therefore, up to numerical accuracies, the Borel resummation coincides with the solution of (\ref{eq:ly:LYeq}).

The question again is if the agreement is trivial.
With that in mind we also plot the matching between the optimal truncation and the integration (which is mostly determined by the precision of the truncation itself).
This serves as a baseline for comparison with the number of matching digits between numeric integration and Borel-Pad\'e. The difference
 tells how much of the agreement is purely due to resummation -- going up to $\sim 40$ digits at small values of $x$.

Finally, one could think that there are exponentially small corrections hidden behind the precisions used.
As detailed in \cite{mar}, the size of such a exponentially small term is precisely that of the error committed by optimal truncation.
These grow as we go to $x\to 0$, and
unless the problem has some fine-tuning of the order of $10^{40}$, we can rule out the need for non-perturbative corrections to the Borel resummation.

	\section{Conclusions and outlook}
The relatively straightforward nature of the problem (as opposed to, say, topological string theories) is at the core of its simple Borel plane structure and well behaved resummation.
The agreement between numerical integration and Borel-Pad\'e is excellent, and no additional non-perturbative corrections are needed.

The missing necessary corrections in \cite{gra-mar-zak} were understood to be coming from complex poles in the positive real part of the Borel plane.
Even though they are present in our case, their corresponding transseries is not needed to complete the Borel resummation.
However, that this kind of equations can be studied numerically
with high precision opens the door to asking the same question about different problems. To begin with, one can think of the other
low-dimensional string equations that are studied in \cite{dif-gin-zin}, along with the Lee-Yang edge singularity.
After finding a case with positive real part instantons, one could ideally use the transseries framework in \cite{cou-mar-sch} to include them
and reproduce exact results.

Of course, one could also study the $x<0$ region here, where the oscillations should be obtained by 
transseries corrections. This lies outside of the purpose of this note, which is studying a case where a \textit{well defined} Borel sum exists,
who can be directly compared to the real numerical answer.

\clearpage

\section*{Acknowledgements}
I would like to thank Marcos Mari\~no for suggesting the problem in the first place, and for his help and advice throughout. I would also
like to thank Szabolcs Zakany for our several discussions on the subject.
This  research  was  supported  in  part  by  the  Fonds  National  Suisse,  subsidies
200021-156995  and  200020-141329,  and  by  the  Swiss-NSF  grant  NCCR  51NF40-141869  “The
Mathematics of Physics” (SwissMAP).

	\clearpage
\section*{Appendix: Numerical tables}
\addcontentsline{toc}{section}{Appendix: Numerical tables}
{\footnotesize All values shown up to their respective precisions.}
\vspace{5ex}
\begin{center}
	\centering
	\tiny
	Optimally truncated $f(x)$\\
	\begin{tabular}{|c|p{9cm}|}
		\hline $x$ & $f(x)$ \\
		\hline $2$ & $1.3$ \\ 
		\hline $4$ & $1.59073$ \\ 
		\hline $6$ & $1.8186377$ \\ 
		\hline $8$ & $2.00086036150$ \\ 
		\hline $10$ & $2.15498728734783$ \\ 
		\hline $12$ & $2.2898129373328068$ \\ 
		\hline $14$ & $2.4104250162655861467$ \\ 
		\hline $16$ & $2.52005872342591954554422$ \\ 
		\hline $18$ & $2.62091262740076192578617567$ \\ 
		\hline $20$ & $2.71455635675233020588851799062$ \\ 
		\hline $22$ & $2.802154016385710992710466847641073$ \\ 
		\hline $24$ & $2.884595523655908604556759014082367622$ \\ 
		\hline $26$ & $2.9625782033935178297395309148386599380309$ \\ 
		\hline $28$ & $3.0366597988814206650903255567321680526928070$ \\ 
		\hline $30$ & $3.10729420863066998154759548851610233619605022702$ \\ 
		\hline $32$ & $3.174856337961121858662584673038457091766057432352291$ \\ 
		\hline $34$ & $3.239659844759811219192249415003574529035280231261226044$ \\ 
		\hline $36$ & $3.3019701041589074417361002236242101151246356402503182361949$ \\ 
		\hline
	\end{tabular}
	\\~\\Numerically integrated $f(x)$, Taylor order $N=60$\\
	\begin{tabular}{|c|p{9cm}|}
		\hline $2$ & $1.2717432154878315125890487374340645$ \\ 
		\hline $4$ & $1.5907325048941121260771523803118030424757552$ \\ 
		\hline $6$ & $1.818637699896956325132568543871547647856616293557$ \\ 
		\hline $8$ & $2.000860361494497989972567430807986033109474607790$ \\ 
		\hline $10$ & $2.1549872873478251122542299372497687073270122470114$ \\ 
		\hline $12$ & $2.28981293733280677038685544831563393614904382518568$ \\ 
		\hline $14$ & $2.410425016265586146739636930437514923208691806928593$ \\ 
		\hline $16$ & $2.520058723425919545544218628825362490985040714408864$ \\ 
		\hline $18$ & $2.6209126274007619257861756675188334307442061180631239$ \\ 
		\hline $20$ & $2.71455635675233020588851799062011322701974934194522064$ \\ 
		\hline $22$ & $2.80215401638571099271046684764107304539723268757593681$ \\ 
		\hline $24$ & $2.884595523655908604556759014082367622249654156156161162$ \\ 
		\hline $26$ & $2.9625782033935178297395309148386599380309061579618881205$ \\ 
		\hline $28$ & $3.03665979888142066509032555673216805269280698175073696155$ \\ 
		\hline $30$ & $3.10729420863066998154759548851610233619605022702175596313$ \\ 
		\hline $32$ & $3.174856337961121858662584673038457091766057432352290860755$ \\ 
		\hline $34$ & $3.2396598447598112191922494150035745290352802312612260439537$ \\ 
		\hline $36$ & $3.30197010415890744173610022362421011512463564025031823619491$ \\ 
		\hline
	\end{tabular}
	\\~\\Borel resummed $f(x)$, Pad\'e order $n=50$\\
	\begin{tabular}{|c|p{9cm}|}
		\hline $2$ & $1.2717432154878315125890$ \\ 
		\hline $4$ & $1.59073250489411212607715238031180304$ \\ 
		\hline $6$ & $1.818637699896956325132568543871547647856616294$ \\ 
		\hline $8$ & $2.00086036149449798997256743080798603310947460778981706$ \\ 
		\hline $10$ & $2.1549872873478251122542299372497687073270122470113775153943730$ \\ 
		\hline $12$ & $2.28981293733280677038685544831563393614904382518568072865361231757313$ \\ 
		\hline $14$ & $2.41042501626558614673963693043751492320869180692859326548946012581739 \newline ~\hspace{1.5em}6092736$ \\ 
		\hline $16$ & $2.52005872342591954554421862882536249098504071440886432329908345089145 \newline ~\hspace{1.5em}3189631457939$ \\ 
		\hline $18$ & $2.62091262740076192578617566751883343074420611806312393173334046725542 \newline ~\hspace{1.5em}3234655503938339219$ \\ 
		\hline $20$ & $2.71455635675233020588851799062011322701974934194522064079741664013737 \newline ~\hspace{1.5em}8811436551557520211509958$ \\ 
		\hline $22$ & $2.80215401638571099271046684764107304539723268757593681299439578930128 \newline ~\hspace{1.5em}742525744665103552883022107125$ \\ 
		\hline $24$ & $2.88459552365590860455675901408236762224965415615616116171479511953545 \newline ~\hspace{1.5em}88403268006388163358122646070823437$ \\ 
		\hline $26$ & $2.96257820339351782973953091483865993803090615796188812055852421329083 \newline ~\hspace{1.5em}7350276265980241275144735302072141389449$ \\ 
		\hline $28$ & $3.03665979888142066509032555673216805269280698175073696154788367706425 \newline ~\hspace{1.5em}718625638609968558625358473003228282943624621$ \\ 
		\hline $30$ & $3.10729420863066998154759548851610233619605022702175596313331762438701 \newline ~\hspace{1.5em}0184876075226167662195168940693240487074795615680$ \\ 
		\hline $32$ & $3.17485633796112185866258467303845709176605743235229086075513405759299 \newline ~\hspace{1.5em}096811920837484471569009836739646200421876561075101448$ \\ 
		\hline $34$ & $3.23965984475981121919224941500357452903528023126122604395369401352933 \newline ~\hspace{1.5em}3761213421861391431387238741733627284129848845870314366538$ \\ 
		\hline $36$ & $3.30197010415890744173610022362421011512463564025031823619490681216661 \newline ~\hspace{1.5em}71740342840579502160123341431627994365596629104943395672639467$ \\ 
		\hline
	\end{tabular}
\end{center}
\end{document}